\documentclass[12pt,sort&compress]{article}

\usepackage{xcolor}

\usepackage{amsmath}
\usepackage{graphicx}
\usepackage{tabularx}
\usepackage{subcaption}
\usepackage{amssymb}
\usepackage{mathtools}
\usepackage{bbm}
\usepackage{lmodern}
\usepackage[sort&compress,numbers]{natbib}
\usepackage{multicol}
\usepackage{multirow}
\usepackage{tikz}
\usepackage{hyperref}
\usepackage{booktabs}
\usepackage[title,titletoc]{appendix}

\usepackage{tabularx,ragged2e}
\newcolumntype{C}{>{\Centering\arraybackslash}X} 
\usepackage[outercaption]{sidecap}    
\usepackage{hyperref}

\title{Motility-Induced Phase Separation  of Run-and-Tumble Particles under Local Stochastic Resetting}

\author{Pascal Grange\\[0.5em]
\small Division of Natural and Applied Sciences, Duke Kunshan University, Kunshan, Jiangsu 215316, China\\
\small Zu Chongzhi Center for Mathematics, Duke Kunshan University, China\\
\small\texttt{pascal.grange@dukekunshan.edu.cn}}

\date{}

\begin{document}
\maketitle

\begin{abstract}
We investigate the non-equilibrium stationary states of a one-dimensional active gas composed of run-and-tumble particles (RTPs) undergoing motility-induced phase separation (MIPS) and subjected to local stochastic resetting at the origin. Unlike global resetting, local resetting maintains a mass-injection flux at the origin while preserving large-scale spatial self-organization. We demonstrate that above a critical resetting rate $r_c$, the active system undergoes a transition toward a stationary two-phase state featuring a dense liquid droplet centered at the origin coexisting with a surrounding gas phase. Using kinetic evolution equations, we derive the exact steady-state density profile expressed in terms of the Lambert function. 
The size of the droplet depends non-monotonically on the resetting rate \(r\), reaching a maximum at an optimal resetting rate.
Furthermore, we analyze the dynamical stability of the supersaturated gas phase and show that a nucleated droplet relaxes exponentially toward its steady-state size with a relaxation timescale of $1/r$. We extend our framework to model a finite resetting velocity, proving that the stationary droplet size is invariant with respect to the resetting velocity.
\end{abstract}

\newpage
\tableofcontents

\section{Introduction}

Stochastic resetting has attracted widespread interest in non-equilibrium statistical physics.  Inspired by optimal target search protocols, resetting induces non-equilibrium steady states with exactly computable spatial distributions and optimizing first-passage properties, first studied for Brownian motion \cite{evans2011diffusion,evans2011optimal,evans2014diffusion} (for a review, see \cite{stochasticReview}). Important extensions to other systems include the study of run-and-tumble particles (RTPs) under resetting.  Run-and-tumble particles provide a model of biological swimmers, such as \emph{Escherichia coli} bacteria \cite{berg1972chemotaxis,schnitzer1993theory,
  berg2004coli, bearon2000modelling, othmer1988models, kurzthaler2024characterization}. In one spatial dimension, exact results have been established for the survival probability, stationary profiles, and first-passage properties of RTPs under various external potentials, boundary conditions, and resetting protocols \cite{malakar2018steady, de2021survival, dhar2019run, angelani2015run, angelani2017confined, angelani2023one, singh2020run, singh2022extremal, bressloff2022encounter, bressloff2023encounter, sarmiento2025first, gueneau2024relating,grange2025mean,grange2026first}.\\

When stochastic resetting is extended to many-body interacting systems,  a fundamental  distinction emerges between \emph{global resetting} and \emph{local resetting} (for a review of stochastic resetting in interacting particle systems, see  \cite{gupta2022stochastic}). Global resetting simultaneously restores the entire system to a predetermined state, instantaneously destroying all pre-existing spatial correlations and  phase structures. Conversely, \emph{local resetting} acts independently on individual particles. In the case of resetting of the position of the constituents to a given point, local resetting establishes a localized 
mass-injection flux without destroying large-scale spatial coherence.
Global resetting has been studied for the Ising model \cite{magoni2020ising}, exclusion processes \cite{basu2019symmetric,sadekar2020zero,karthika2020totally}, as well as predator-prey models \cite{toledo2019predator,evans2022exactly,mercado2018lotka}, fluctuating interfaces \cite{gupta2014fluctuating,gupta2016resetting}, synchronization \cite{sarkar2022synchronization},  reaction-diffusion processes \cite{durang2014statistical}, and glassy systems \cite{grange2020entropy}.
Stationary states under local resetting have been explicitly calculated in models of binary aggregation \cite{grange2021aggregation}  and exclusion processes  \cite{miron2021diffusion,pelizzola2021simple}, as well as in the dynamics of spin models and zero-range processes \cite{grange2022voter,grange2019steady,grange2020non,grange2024local}, and population-growth models with quenched growth heterogeneity \cite{bernard2026mean,bernard2026partial}. For resetting protocols acting on subsystems of systems of interacting spins or coupled oscillators, see \cite{acharya2025manipulating}\\

 On the other hand, motility-induced phase separation (MIPS) has established itself as a foundational principle of active-matter physics \cite{tailleur2008statistical, cates2015motility, thompson2011lattice, ramaswamy2010mechanics}. Driven by a density-dependent slowdown $v(\rho)$, active particles may undergo phase separation into a dense (liquid) phase and a dilute (gas) phase in the absence of attractive interactions. However, how MIPS mechanisms compete with or adapt to local stochastic resetting remains an open question. In this paper, we establish the exact one-dimensional phase diagram of a system of RTPs with a density-dependent slowdown in the presence of local Poissonian resetting to the origin at rate $r$. Intuitively, the resetting prescription creates a persistent flux of matter at the origin, driving up the density around the origin. If the resetting rate is strong enough, a stationary liquid droplet centered at the origin may emerge, surrounded by a gas phase.\\

The paper is organized as follows. In Section~\ref{sec:evolution}, we present the kinetic evolution equations and derive a quantity conserved in each phase as a first integral. In Section~\ref{sec:droplet}, we construct the exact stationary biphase solution, determine the density profiles and study the width of the liquid droplet surrounding the origin as a function of the resetting rate. Section~\ref{sec:gas} analyzes the single gas phase solution, supersaturation, and droplet-nucleation dynamics. Section~\ref{sec:finite_velocity} extends the model to finite resetting velocities.

\section{Evolution equations and first integral}
\label{sec:evolution}

We consider a system of RTPs moving on the real line. Each particle performs runs at velocity $\pm v(\rho)$ interspersed with instantaneous tumbles, where $\rho$ is the local density of particles. The tumbles occur at random times, distributed according to a Poisson process of intensity $\alpha$. Upon tumbling, the particle uniformly picks its direction of motion: it reverses its direction of motion with probability $1/2$ (and keeps unchanged\footnote{one could equivalently consider a Poisson process of intensity $\alpha/2$, and declare that the RTP flips the direction of its motion with probability $1$ at tumble events. The description with intensity $\alpha$ generalizes to higher dimension, where the RTP uniformly picks its direction of motion at each tumble event.} with probability $1/2$). To incorporate motility-induced phase separation, the running velocity $v(\rho)$ is taken as a decreasing function of the local density $\rho(x,t)$, following the piecewise prescription introduced in \cite{thompson2011lattice}:
\begin{equation}\label{eq:vrho}
v(\rho) = v_0\left( 1 - \frac{\rho}{\rho_m} \right)\mathbbm{1}( \rho < \rho_m) + \frac{v_0}{\rho_m}\mathbbm{1}( \rho \geq \rho_m),
\end{equation}
where $v_0$ is the free-particle velocity and $\rho_m$ denotes a characteristic density parameter dividing the system into a dense liquid phase ($\rho \geq \rho_m$), and a gas phase ($\rho < \rho_m$) with a density-dependent slowdown.\\

 Particles running at velocity $+v(\rho)$ (resp. $-v(\rho)$) are termed right-moving (resp. left-moving).  
Let $\rho_+(x,t)$ and $\rho_-(x,t)$ denote the spatial densities of right-moving and left-moving particles at position $x$ and time $t$, respectively.
 The total  density $\rho$ is the sum of the densities of left-movers and right-movers: 
\begin{equation}
\rho(x,t) = \rho_+(x,t) + \rho_-(x,t).
\end{equation}

Each particle is independently reset to the origin at times distributed according to a Poisson process of intensity $r$ (called the resetting rate). For $x \neq 0$, the kinetic equations governing the density fields are the following modified telegrapher's equations:
\begin{equation}\label{eq:master}
\begin{split}
\frac{\partial \rho_+}{\partial t} &= -\frac{\partial}{\partial x}\left( v(\rho) \rho_+  \right) - \frac{\alpha}{2} \rho_+ +\frac{\alpha}{2}\rho_- - r \rho_+,\\
\frac{\partial \rho_-}{\partial t} &= \frac{\partial}{\partial x}\left( v(\rho) \rho_-  \right) - \frac{\alpha}{2} \rho_- +\frac{\alpha}{2}\rho_+ - r \rho_-.
\end{split}    
\end{equation}
 The values $\rho_\pm(0,t)$ of the densities at the origin  are such that the constraint of mass conservation is satisfied:
\begin{equation}\label{eq:massCons}
 M = \int_{-\infty}^{+\infty} \left( \rho_+(x,t) + \rho_-(x,t)\right) dx,
\end{equation}
 where $M$ is the total mass of the system. The parameters of the models are therefore $v_0,\alpha,\rho_m, r$ and $M$.\\

In the liquid phase ($\rho \ge \rho_m$), the absolute value of the velocity is constant ($v(\rho) = v_0/\rho_m$), and Eqs.~\eqref{eq:master} reduce to a linear system:
\begin{equation}\label{eq:evolEqLiquid}
\begin{split}
\frac{\partial \rho_+}{\partial t} &= -\frac{v_0}{\rho_m} \frac{\partial \rho_+}{\partial x} - \left(r + \frac{\alpha}{2}\right) \rho_+ + \frac{\alpha}{2}\rho_-,\\
\frac{\partial \rho_-}{\partial t} &= \frac{v_0}{\rho_m} \frac{\partial \rho_-}{\partial x} - \left(r + \frac{\alpha}{2}\right) \rho_- + \frac{\alpha}{2}\rho_+, \qquad (\rho \ge \rho_m).
\end{split}    
\end{equation}
In the gas phase ($\rho < \rho_m$), substituting $v(\rho) = v_0 (1 - \rho/\rho_m)$ yields a non-linear system:
\begin{equation}\label{eq:evolEqGas}
\begin{split}
\frac{\partial \rho_+}{\partial t} &= -v_0\frac{\partial \rho_+}{\partial x} + \frac{v_0(\rho_+ + \rho_-)}{\rho_m} \frac{\partial \rho_+}{\partial x} + \frac{v_0\rho_+}{\rho_m} \frac{\partial (\rho_+ + \rho_-)}{\partial x} - \left(r + \frac{\alpha}{2}\right) \rho_+ + \frac{\alpha}{2}\rho_-,\\
\frac{\partial \rho_-}{\partial t} &= v_0 \frac{\partial \rho_-}{\partial x} - \frac{v_0(\rho_+ + \rho_-)}{\rho_m} \frac{\partial \rho_-}{\partial x} - \frac{v_0\rho_-}{\rho_m} \frac{\partial (\rho_+ + \rho_-)}{\partial x} - \left(r + \frac{\alpha}{2}\right) \rho_- + \frac{\alpha}{2}\rho_+, \qquad (\rho < \rho_m).
\end{split}    
\end{equation}

To analyze stationary states, we look for solutions that do not depend on time,  denoted by $\rho_+(x)$, $\rho_-(x)$.
 Let us introduce the total stationary density and total current 
 \begin{equation}
 \begin{split}
     \rho(x):=& \rho_+(x) + \rho_-(x),\\
     \sigma(x) :=& \rho_+(x) - \rho_-(x),
     \end{split}
 \end{equation}
 and define two modified fluxes:
\begin{equation}\label{defRS}
\begin{split}
R(x) :=& v(\rho(x)) \rho(x),\\
 S(x) :=& v(\rho(x)) \sigma(x).
\end{split}
\end{equation}
In the stationary state, the sum and difference of the equations in Eq. \eqref{eq:evolEqGas} yield the following system of equations:
\begin{equation}\label{eq:stationary_RS}
\begin{split}
0 &= \frac{d S(x)}{d x} + r\rho(x),\\
0 &= \frac{d R(x)}{d x} + (r+\alpha)\sigma(x).
\end{split}    
\end{equation}
Multiplying the first equation by $(r+\alpha) v(\rho(x)) S(x)$ and the second by $r v(\rho(x)) R(x)$, and taking their difference, we obtain:
\begin{equation}
(r+\alpha) S(x) \frac{d S(x)}{d x} - r R(x) \frac{d R(x)}{d x} = 0.
\end{equation}
Integrating on an interval contained in  a continuous phase of the system demonstrates that there exists a constant $K$  in the phase such that:
\begin{equation}\label{eq:conservation}
S(x)^2 - \frac{r}{r+\alpha} R(x)^2 = K.
\end{equation}

\section{Steady state with two phases}
\label{sec:droplet}

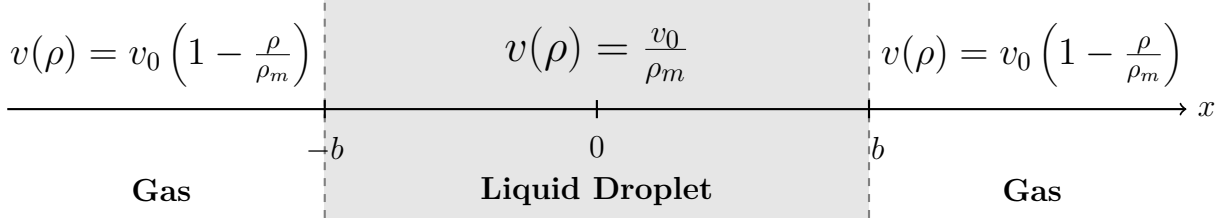
\begin{figure}[htbp]
    \centering
    \begin{tikzpicture}[scale=1.2]
        \fill[black!10] (-3, -1.2) rectangle (3, 1.2);
        
        \draw[dashed, thick, black!50] (-3, -1.2) -- (-3, 1.2);
        \draw[dashed, thick, black!50] (3, -1.2) -- (3, 1.2);
        
        \draw[->, thick] (-6.5, 0) -- (6.5, 0) node[right] {$x$};
        
        \draw[thick] (0, 0.1) -- (0, -0.1) node[below=0.1cm] {$0$};
        \draw[thick] (-3, 0.1) -- (-3, -0.1) node[below=0.1cm] {$-b$};
        \draw[thick] (3, 0.1) -- (3, -0.1) node[below=0.1cm] {$\;\;b$};
        
        \node at (0, 0.6) {\Large $v(\rho) = \frac{v_0}{\rho_m}$};
        \node at (0, -0.9) {\textbf{Liquid Droplet}};
        
        \node at (-4.8, 0.6) {\large $v(\rho) = v_0\left( 1 - \frac{\rho}{\rho_m}\right)$};
        \node at (-4.8, -0.9) {\textbf{Gas}};
        
        \node at (4.8, 0.6) {\large $v(\rho) = v_0\left( 1 - \frac{\rho}{\rho_m}\right)$};
        \node at (4.8, -0.9) {\textbf{Gas}};
    \end{tikzpicture}
    \caption{Spatial configuration of the steady state with a central liquid droplet of half-width $b$. The particle velocity $v(\rho)$ switches from a density-dependent profile in the gas phase to a uniform minimum value inside the saturated liquid phase.}
    \label{fig:schema_droplet}
\end{figure}

Let us search for a  non-equilibrium steady state consisting of a liquid droplet occupying the central domain $[-b, b]$, bounded by a gas phase occupying the domain of equation $|x| > b$. Intuitively, such a state can exist if the resetting is strong enough to sustain a dense phase around the origin.  
 Such a configuration is presented schematically in Fig. \ref{fig:schema_droplet}.  The system is parity invariant. Hence,
\begin{equation}
 \rho_{\pm}(-x) = \rho_\mp( -x),\qquad (x\in \mathbbm{R}).
\end{equation}
We therefore restrict the discussion to positive values of the coordinate $x$.\\

\subsection{Interface equilibrium}

The liquid-gas interface $x = b$ is the point where the liquid reaches the minimum value compatible with the liquid  state. Moreover, mechanical stability requires continuity of the flux of matter. Hence, 
\begin{equation}\label{eq:equilInterface}
 \begin{split}   
    \rho(b^-) =& \rho_m,\\
    \frac{v_0}{\rho_m} \rho(b^-) =& v_0\left( 1 - \frac{\rho(b^+)}{\rho_m}\right)\rho(b^+).
\end{split}
\end{equation} 
 Combining these two conditions yields
\begin{equation}\label{quadrhobplus}
1 = \left(1 - \frac{\rho(b^+)}{\rho_m}\right) \rho(b^+),
\end{equation}
The r.h.s. is a quadratic function of $\rho(b^+)$ attaining its maximum value $\rho_m/4$ at $\rho(b^+) = \rho_m/2$.
For a steady state with two phases to exist, the value $1$ of the l.h.s. must be in the interval $[0,\rho_m/4]$. The density parameter $\rho_m$ must therefore satisfy
\begin{equation}
\rho_m \geq 4.
\end{equation}
Assuming this condition is satisfied,
we obtain the gas density at the interface:
\begin{equation}\label{eq:rhobplus}
\rho(b^+) = \frac{\rho_m}{2}\left( 1 - \sqrt{1 - \frac{4}{\rho_m}} \right).
\end{equation}

In the far-field gas region ($|x| \to \infty$), the particle densities $\rho_+(x)$ and $\rho_-(x)$ decay to zero, enforcing $K = 0$ in Eq.~\eqref{eq:conservation}. Hence, 
\begin{equation}\label{carro}
S(x)^2 = \frac{r}{r+\alpha} R(x)^2.
\end{equation}
As the total density is positive, $R(x)$ is also positive. On the other hand, we know from Eq.  \eqref{eq:stationary_RS} that $S'(x) = -r\rho(x)$ so $S(x)$ is a decreasing function. As it goes to zero when $x$ approaches infinity, $S(x)$ and $\sigma(x)$ must be positive. We can therefore take the square root of both sides of Eq. \eqref{carro} 
 to establish the proportionality relation between current and density throughout the gas phase:
\begin{equation}\label{eq:propAbs}
S(x) = \sqrt{\frac{r}{r+\alpha}} R(x), \qquad 
 {\mathrm{hence}}\qquad \sigma(x) = \sqrt{\frac{r}{r+\alpha}} \rho(x),\qquad(|x|>b).
\end{equation}
In the domain $[0,b]$, substituting $v(\rho) = v_0/\rho_m$ into Eq. \eqref{eq:stationary_RS}  yields the differential equation
\begin{equation}
    \frac{d^2\sigma}{dx^2} = - \frac{\rho_m^2 r(r+\alpha)}{v_0^2} \sigma.
\end{equation}
Hence, the current $\sigma(x)$ is a linear combination of exponential functions $\exp(\pm \rho_m \chi x)$ of the form
\begin{equation}
\sigma(x) = \sigma_+ e^{ \rho_m \chi x} + \sigma_- e^{ -\rho_m \chi x},\qquad(x\in[0,b]),
\end{equation}
with the following notation for the characteristic inverse length scale: 
\begin{equation}\label{eq:chi_def}
\chi := \frac{1}{v_0}\sqrt{r(\alpha + r)}.
\end{equation}
 Using Eq. (\ref{eq:stationary_RS}) yields $\rho(x) = - v_0 \chi r^{-1}\sigma'(x)$. 
 By continuity of the currents $v(\rho)\rho_\pm$ on $[0,+\infty[$, the first integral is
 $K=0$ and $\sigma_+^l=0$, hence for $x$ in $[0,b]$,
\begin{equation}
\sigma( x ) = 
  \sigma_-^l e^{-\rho_m \chi x},\;\;\;\;\rho(x) = \frac{v_0 \chi}{r}\sigma_-^l e^{-\rho_m \chi x}. 
\end{equation}
In particular, using Eq. (\ref{eq:equilInterface}), 
\begin{equation}
\rho_m = \frac{v_0 \chi}{r}\sigma_-^l e^{-\rho_m \chi b}. 
\end{equation}

Using parity completely specifies the density and current profiles in the liquid phase:  
\begin{equation}\label{eq:liquid_profiles}
\begin{split}
\rho(x) &= \rho_m e^{-\rho_m \chi (x - b)},\\
\sigma(x) &= \rho_m \frac{r}{v_0 \chi} e^{-\rho_m \chi (x - b)}, \qquad (0 \leq x \leq b).
\end{split}
\end{equation}
Integrating the density $\rho$ over $[-b, b]$ yields the total mass $M_{\text{liq}}$ contained in the liquid droplet:
\begin{equation}\label{eq:Mliq}
M_{\text{liq}} := \int_{-b}^b \rho(x) dx = \frac{2}{\chi}\left( e^{\rho_m \chi b} - 1 \right).
\end{equation}

\subsection{Density profile in the gas phase}

Consider the gas phase occupying the region $]b,+\infty[$). Substituting the proportionality relation between $\sigma(x)$ and $\rho(x)$ obtained in Eq. \eqref{eq:propAbs} into Eq.~\eqref{eq:stationary_RS}  gives:
\begin{equation}
\frac{d}{dx}\left( v_0\left(  1 - \frac{\rho(x)}{\rho_m} \right) \rho(x) \right)+ \sqrt{r(r+\alpha)} \rho(x)=0.  
\end{equation}
The total density $\rho$ in the gas phase therefore satisfies the separable differential equation
\begin{equation}\label{eq:separable}
\rho'(x) - \frac{2\rho(x)\rho'(x)}{\rho_m} = -\chi \rho(x).
\end{equation}
Separating variables and integrating from $b$ to $x>b$ yields:
\begin{equation}\label{preSep}
\ln\left( \frac{\rho(x)}{\rho(b^+)} \right) - \frac{2}{\rho_m}\left( \rho(x) - \rho(b^+) \right) = -\chi(x - b).
\end{equation}
Exponentiating yields an implicit solution:
\begin{equation}\label{preSep}
\rho(x) \exp\left( -\frac{2}{\rho_m} \rho(x) \right) = 
\rho(b^+) \exp\left( -\frac{2}{\rho_m} \rho(b^+) -\chi(x - b) \right).
\end{equation}
To invert this relation, we need the principal branch $W_0$ of the Lambert $W$ function, defined on $[-e^{-1},+\infty[$ by the relation $W_0(z) e^{W_0(z)}=z$  (see Eq. (4.13.1) in \cite{nist_dlmf}). The  gas density profile is expressed as  
\begin{equation}\label{eq:rhogas_lambert}
\rho(x) = -\frac{\rho_m}{2} W_0\left( -\frac{2\rho(b^+)}{\rho_m} \exp\left[ -\frac{2\rho(b^+)}{\rho_m} -\chi (x-b) \right] \right), \qquad (x > b),
\end{equation}
 where $\rho(b^+)$ has been obtained in Eq. (\eqref{eq:rhobplus}) as an equilibrium condition for the interface. As we have seen from Eq. \eqref{quadrhobplus} that $\rho(b^+)\leq \rho_m/2$, the argument of the Lambert function is in the interval $[-e^{-1},0]$, which justifies the use of the principal branch $W_0$.
 The Lambert function also occurs in the stationary state of exclusion processes with boundary conditions studied in the thermodynamic limit in \cite{pelizzola2021simple}, where  evolution equations containing non-linearities in the resetting terms are solved in the mean-field approximation.\\ 

To determine the half-width $b$ of the droplet, we enforce mass conservation. By parity symmetry, the half-space $[0, \infty)$ must contain half of the mass of the system in the steady state. Eq. (\ref{eq:massCons}) therefore yields
\begin{equation}\label{eq:mass_conservation_integral}
\begin{split}
&\frac{M}{2} = \int_0^b \rho_m e^{-\rho_m \chi (x-b)} dx + \int_b^\infty -\frac{\rho_m}{2} W_0\left( -A e^{-\chi(x-b)} \right) dx,\\
&\text{with}\qquad A = \frac{2\rho(b^+)}{\rho_m} e^{-2\rho(b^+)/\rho_m}.
\end{split}
\end{equation}
 Introducing the variable $z := W_0\left(-A e^{-x}\right)$ simplifies the second integral in the r.h.s. as follows. Indeed, using the definition of the Lambert function and differentiating,
\begin{equation}\label{changeVar}
\begin{split}
ze^{z} =& -A e^{-x}\\ 
( 1 + z)e^{z} dz =& A e^{-x} dx\\
( 1 + z)e^{z} dz =& -ze^{z} dx\\
z\,dx =& -( 1 + z) dz.\\
\end{split}
\end{equation}
Hence, the mass of the gas phase occupying the domain $[b,+\infty[$ is expressed as 
\begin{equation}
\int_b^\infty \rho(x) dx = -\frac{\rho_m}{2\chi} \int_{-2\rho(b^+)/\rho_m}^0 (1 + z) dz = \frac{\rho_m}{2\chi} \left[ \frac{2\rho(b^+)}{\rho_m} - \frac{1}{2}\left(\frac{2\rho(b^+)}{\rho_m}\right)^2 \right] = \frac{1}{\chi},
\end{equation}
where we used Eq. \eqref{quadrhobplus} satisfied by the density $\rho(b^+)$.\\

Substituting this expression back into Eq.~\eqref{eq:mass_conservation_integral} simplifies mass conservation to:
\begin{equation}
\frac{M}{2} = \frac{1}{\chi}( e^{\rho_m \chi b} - 1) + \frac{1}{\chi} = \frac{1}{\chi} e^{\rho_m \chi b}.
\end{equation}
Solving for $b$ yields the exact expression for the droplet half-width:
\begin{equation}\label{eq:droplet_width_exact}
b = \frac{1}{\rho_m \chi}\log\left( \frac{M\chi}{2} \right) = \frac{v_0}{\rho_m\sqrt{r(r+\alpha)}}\log\left( \frac{M\sqrt{r(r+\alpha)}}{2 v_0} \right).
\end{equation}
 This quantity represents the half-width of a droplet only if it is positive. Hence the condition of existence of a stationary state with two phases:
 \begin{equation}\label{condEx}
   M> \frac{\sqrt{r(r+\alpha)}}{2v_0}.
 \end{equation}

\subsection{Phase diagram and non-monotonic droplet  size}

 At fixed total mass $M$, Eq. \eqref{condEx} implies that a droplet exists in the stationary state if the resetting rate is above the critical value $r_c$ defined by 
\begin{equation}\label{eq:rc}
r_c := \frac{\alpha}{2}\left( -1 + \sqrt{1 + \frac{16 v_0^2}{\alpha^2 M^2}} \right).
\end{equation}
For $r < r_c$, the resetting is not frequent enough to maintain localized condensation, and the system remains in a purely gaseous state.\\

\begin{figure}[htbp]
\centering
\includegraphics[width=0.75\linewidth]{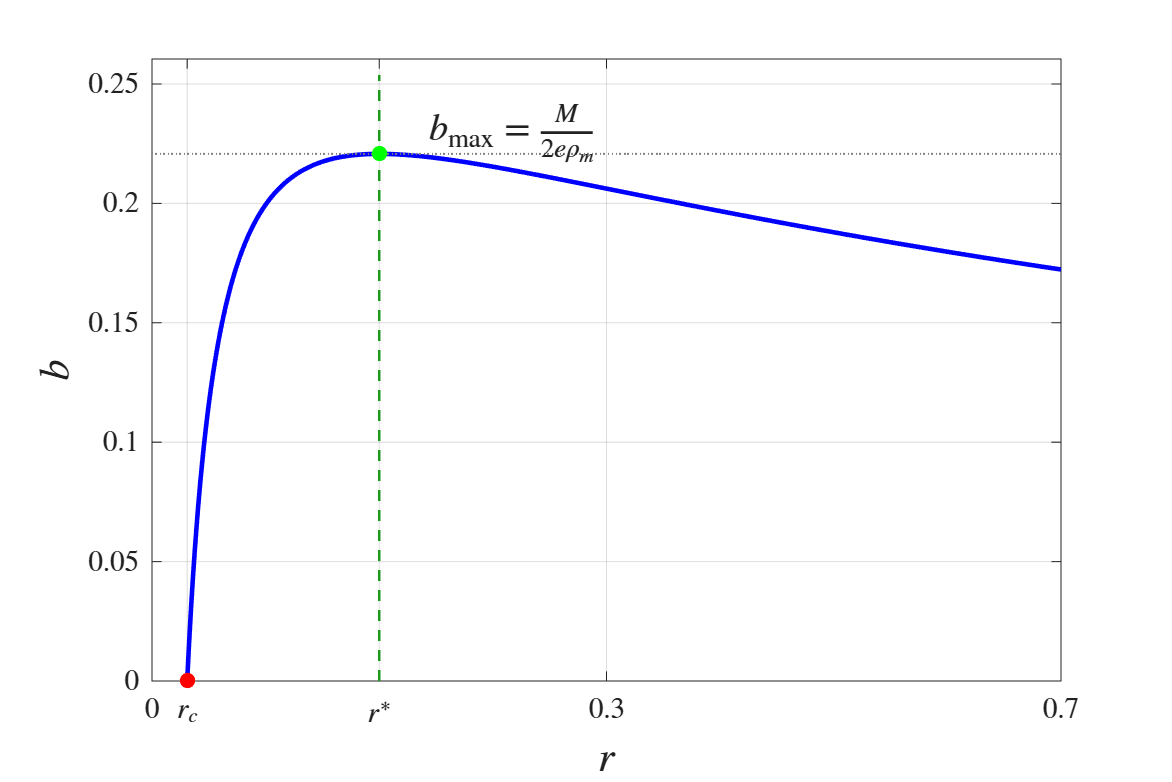}
\caption{Exact droplet half-width $b$ as a function of resetting rate $r$. The droplet emerges at $r_c$, reaches maximum extent at $r^*$, and contracts asymptotically for large $r$.}
\label{fig:figureWidth}
\end{figure}

Moreover, differentiating $b$ with respect to $r$ reveals a non-monotonic dependence of the droplet size on the resetting rate:
\begin{equation}
\frac{\partial b}{\partial r} = \frac{\partial b}{\partial \chi} \frac{\partial \chi}{\partial r} = \frac{1}{\rho_m \chi^2}\left[ 1 - \log\left( \frac{M\chi}{2} \right) \right] \frac{\partial \chi}{\partial r}.
\end{equation}
Setting $\partial b / \partial r = 0$ yields an optimal rate $r^* > r_c$ at which the droplet width reaches its maximum:
\begin{equation}\label{eq:ropt}
r^\ast = \frac{\alpha}{2}\left( -1 + \sqrt{1 + \frac{16 e^2 v_0^2}{\alpha^2 M^2}} \right).
\end{equation}
Intuitively, for $r_c < r < r^\ast$, increasing $r$ enhances the  flux of particles toward the origin, expanding the droplet. For $r > r^\ast$, strong resetting confines particles to ballistic runs near the origin before they can reach the droplet boundary, causing the width of the droplet to shrink. The length $b$ is plotted as a function of the resetting rate on Fig. \ref{fig:figureWidth}.\\ 

On the other hand, the total liquid mass and average liquid density $\langle \rho \rangle_{\text{liq}} = M_{\text{liq}} / (2b)$ are expressed using  Eqs (\ref{eq:Mliq},\ref{eq:droplet_width_exact}) as
\begin{equation}\label{eq:MassDensityLiq}
M_{\text{liq}} = M - \frac{2 v_0}{\sqrt{r(r+\alpha)}}, \qquad \langle \rho \rangle_{\text{liq}} = \rho_m \frac{\frac{M\sqrt{r(r+\alpha)}}{2 v_0} - 1}{\log\left( \frac{M\sqrt{r(r+\alpha)}}{2 v_0}\right)}.
\end{equation}
Both $M_{\text{liq}}$ and $\langle \rho \rangle_{\text{liq}}$ are monotonically increasing functions of $r$ for all $r > r_c$.

\begin{figure}[htbp]
\centering
\includegraphics[width=0.85\linewidth]{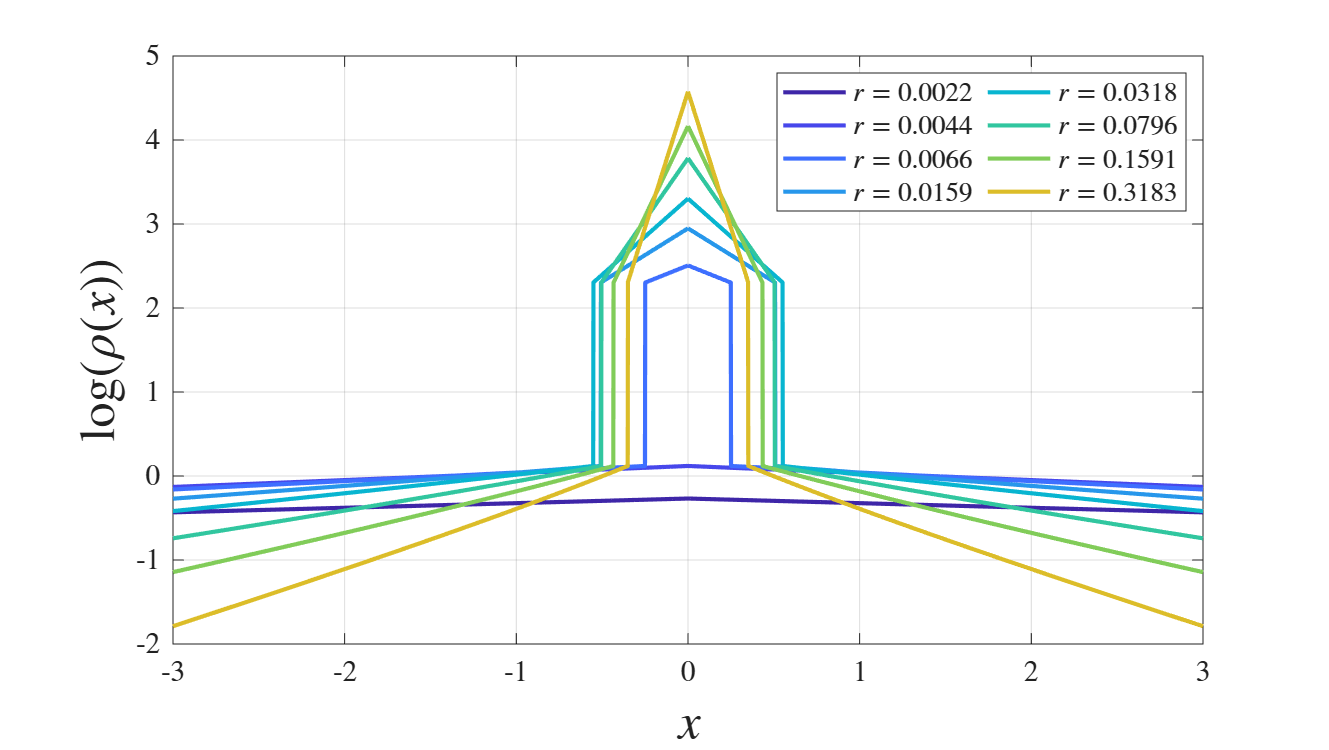}
\caption{Stationary spatial density profile $\rho(x)$ (logarithmic plot) for representative parameters (for numerical purposes, $\alpha =1$, $v_0=1$, $\rho_m=10$ and $M=30$, so that $r_c \simeq 0.0044$ and $r^\ast  \simeq 0.0318$). For $r<r_c$ there is a discontinuity at $x=\pm b$, where $b$ is the half-width of the liquid droplet surrounding the origin (which is maximum for $r=r^\ast$).}
\label{fig:figureDensity}
\end{figure}

\section{Single phase and supersaturation}\label{sec:gas}

\subsection{Gas steady state}

 We know that a steady state with two phases exists only if $\rho_m>4$ and the resetting rate is above the critical value $r_c$ expressed in Eq. \eqref{eq:rc}. A solution with a single liquid phase cannot exist on the entire real line for a system with finite mass because the density of the liquid is bounded from below by 
  $\rho_m$. The other possible steady state therefore has a single (gas) phase.\\

If we seek a single-phase gas solution across the  entire real line, we have to integrate Eq.~\eqref{eq:separable}. As this equation is separable, we can integrate it on $[0,x]$ with $x>0$
  repeating the reasoning of  Eqs (\ref{preSep},\ref{eq:rhogas_lambert}), which yields
\begin{equation}\label{gasProfile}
\rho(x) = -\frac{\rho_m}{2}W_0\left( -\frac{2\rho(0)}{\rho_m} e^{ -\frac{2\rho(0)}{\rho_m} -\chi x}  \right), \qquad (x > 0).
\end{equation}
Parity symmetry and  mass conservation imply
\begin{equation}
\frac{M}{2} = \int_0^\infty \rho(x) dx.
\end{equation}
 Using the same change of variables as in Eqs \eqref{changeVar}
yields a quadratic equation for the value of the density at the origin:
\begin{equation}\label{peakDensity}
\frac{\rho(0)^2}{\rho_m} - \rho(0) + \frac{M\chi}{2} = 0. 
\end{equation}
A pure-gas steady state therefore exists only if 
\begin{equation}\label{condMassGas}
    M\chi \le \rho_m / 2.
\end{equation}
 If this condition is satisfied, 
\begin{equation}
\rho(0) = \frac{\rho_m}{2}\left( 1 - \sqrt{1 - \frac{2M\chi}{\rho_m} } \right).
\end{equation}
Substituting into Eq. \eqref{gasProfile}, the density profile in the system with a single gas phase is expressed in the entire space as
\begin{equation}\label{gasProfile}
\rho(x ) = - \frac{\rho_m}{2}
W_0\left( \left( -1 + \sqrt{1 - \frac{2M\sqrt{r(r+\alpha)}}{v_0\rho_m}}      \right) e^{  -1 + \sqrt{1 - \frac{2M\sqrt{r(r+\alpha)}}{v_0\rho_m}} - 
 \frac{\sqrt{r(r+\alpha)}}{v_0}|x|}   
 \right), \qquad (x\in \mathbb{R}).
\end{equation}
The density profiles $\rho$ obtained in Eqs (\ref{eq:liquid_profiles},\ref{eq:rhogas_lambert},\ref{gasProfile})
 are plotted in Fig. \ref{fig:figureDensity}.

\subsection{Gas metastability and droplet nucleation dynamics}

The condition we have obtained in Eq. \eqref{condMassGas} is compatible with the one obtained in Eq. \eqref{condEx}, if $\rho_m \geq 4$. Hence, if $\rho_m$ is large enough for a steady  state with two phases to exist, the following condition is satisfied for some values of the mass:
\begin{equation}\label{condMassSuper}
   2 < \frac{M\sqrt{r(r+\alpha)}}{v_0} < \frac{\rho_m}{2}.
\end{equation}
 Intuitively, in such a situation the single-phase steady state corresponds to a supersaturated gas. Indeed the resetting is strong enough (at constant mass), or the mass is large enough (at constant resetting) for a liquid phase to exist around the origin, but there exists another steady state consisting of a single gas phase. 
 This can be confirmed by studying the dynamics of a droplet nucleating at the origin.\\ 

\begin{figure}[htbp]
\centering
\includegraphics[width=0.95\linewidth]{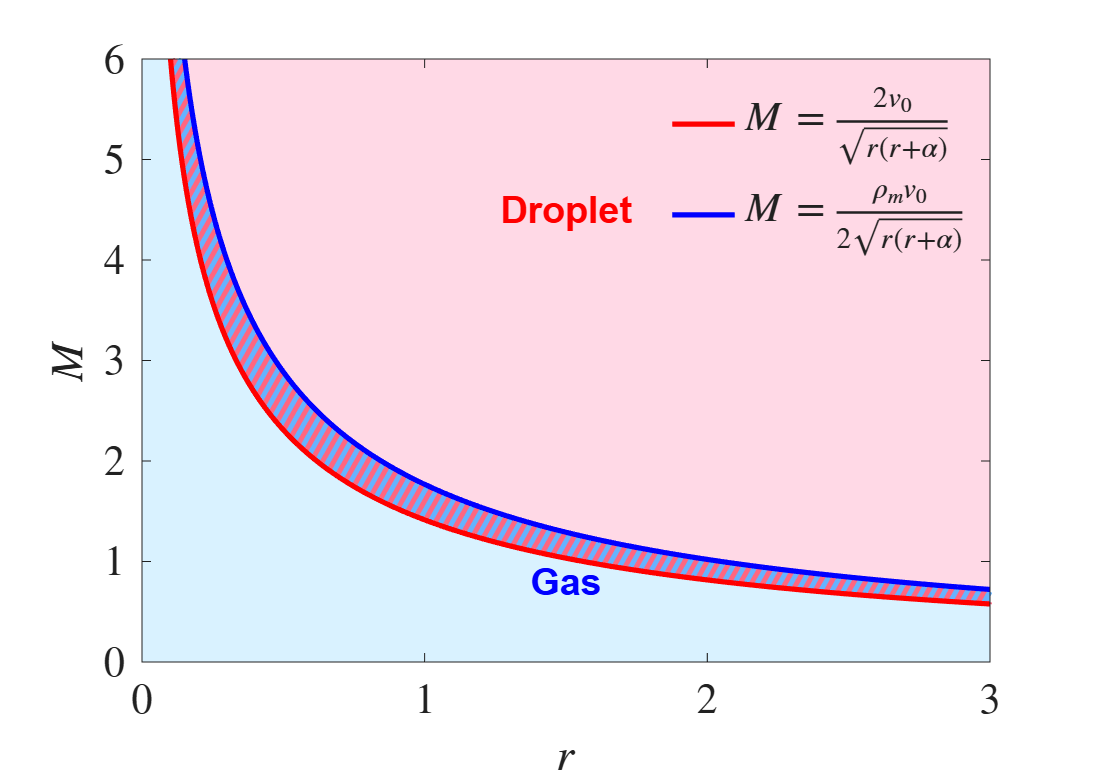}
\caption{Phase diagram. In the domain between the two curves, there are two steady states, one of which is a supersaturated gas.  
 The values used in the plot are $\rho_m =5$, $\alpha = 1$, $v_0=1$.}
\label{fig:phaseDiagram}
\end{figure}

 Let us assume the mass of the system and the resetting rate are such that the condition of Eq. \eqref{condMassSuper} is satisfied. 
Consider a small droplet nucleating at the origin. It occupies the domain $[-l(t),l(t)]$ at time $t$. Let $M_{\text{liq}}(t)$ denote the mass of the droplet at time $t$. This quantity changes at a rate governed by the flux imbalance across its interfaces. By parity symmetry,
\begin{equation}
\frac{1}{2} \frac{d M_{\text{liq}}}{d t} = -\frac{v_0}{\rho_m} \sigma(l(t)^-) + v_0\left( 1 - \frac{\rho(l(t)^+)}{\rho_m}\right)\sigma(l(t)^+).
\end{equation}
Under a quasi-static approximation, this flux balance yields the linear relaxation equation:
\begin{equation}
\frac{1}{2}\frac{d M_{\text{liq}}}{dt} = r\left( \frac{M}{2} - \frac{2v_0}{\sqrt{r(r+\alpha)}}\right) - \frac{r}{2} M_{\text{liq}}(t).
\end{equation}
Integrating with initial condition $M_{\text{liq}}(0) = 0$ yields the explicit growth trajectory:
\begin{equation}\label{eq:nucleation}
M_{\text{liq}}(t) = \left( M - \frac{2 v_0}{\sqrt{r(r+\alpha)}} \right)\left( 1 - e^{-rt} \right).
\end{equation}
 The large-time limit of the mass of the droplet is the mass of the liquid phase calculated in Eq. (\ref{eq:MassDensityLiq}).  
The supersaturated state therefore decays exponentially into the stable two-phase steady state with a characteristic relaxation timescale $1/r$. This prediction  matches direct numerical simulations described in the Appendix (see Fig.~\ref{fig:figureMassDroplet}).  If there are two steady states, the one with a single phase is therefore metastable, and fluctuations lead  to the condensation of a droplet around the origin. The phase diagram is displayed in Fig.  \ref{fig:phaseDiagram}.

\begin{figure}[htbp]
\centering
\includegraphics[width=0.75\linewidth]{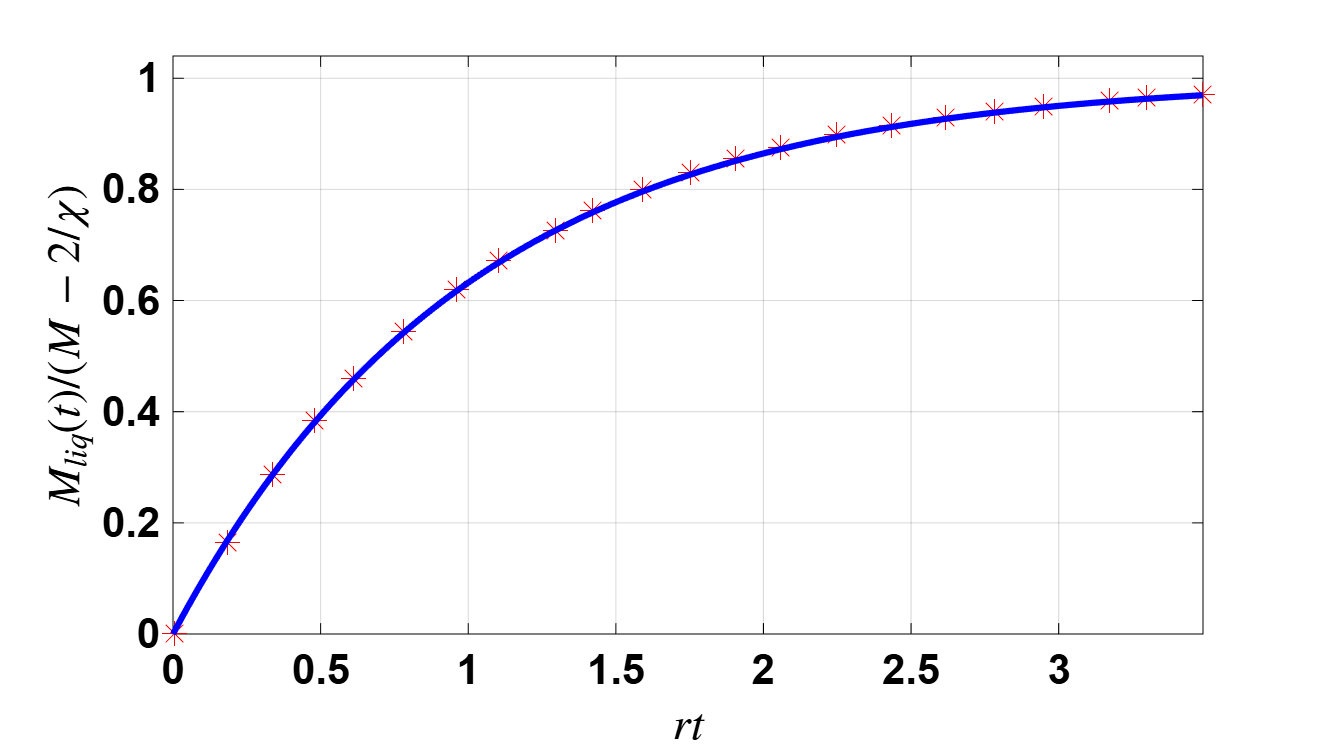}
\caption{Time evolution of droplet mass $M_{\text{liq}}(t)$ nucleating at the origin in a supersaturated gas (for implementation, $v_0=1$, $\alpha = 1$, $\rho_m = 1,000$ $M=289>2/\chi, r=0.5$), confirming exponential relaxation with rate $r$. The continuous curve is the graph of $1-e^{rt}$ Stars indicate the values obtained by direct numerical simulation of the system, as described in Appendix \ref{app:discrete}.}
\label{fig:figureMassDroplet}
\end{figure}

\section{Extension to finite resetting velocity}
\label{sec:finite_velocity}

 The model we have solved assumes that the resetting of  the particles to the origin occurs instantaneously. This model is an idealization of a situation where the particles are running and tumbling on a track, and dropped on a fast-turning treadmill that sends them back to the origin when they undergo resetting. More realistic protocols incorporating finite return velocities, refractory periods, or intermittent confining potentials have been theoretically studied \cite{evans2018effects, mercado2020intermittent,pal2019time, pal2019invariants}. Such non-instantaneous resetting mechanisms have also been realized experimentally \cite{tal2020experimental}, revealing remarkable invariance properties of spatial stationary profiles with respect to the return dynamics.\\

Let us consider a  modification where resetting is non-instantaneous: particles return to the origin at a finite velocity upon resetting, and undergo a density-dependent slowdown on their way to the origin. Particles can be in three states: running with positive velocity (the density of particles in this state is still denoted by $\rho_+$),  running with negative velocity (the density of particles in this state is still denoted by $\rho_-$), and resetting to the origin. Let us denote by $\rho_r(x)$  the density of resetting particles at coordinate $x$ in the steady state. 
The total density  $\rho_{\text{tot}}$ is defined as follows:
\begin{equation}
\begin{split}
    \rho_{\text{tot}}(x) :=& \rho(x) + \rho_r(x),\\
    {\text{with}}  \qquad \rho(x) =& \rho_+(x) + \rho_-(x).,
 \end{split}   
\end{equation}
 Let us assume that resetting particles are traveling to the origin at velocity 
\begin{equation}
v_r(\rho_{\text{tot}}) := \gamma v(\rho_{\text{tot}}),
\end{equation}
where $\gamma > 0$ parametrizes the relative resetting speed. The model with instantaneous resetting is recovered in the limit where $\gamma$ approaches infinity, while the value $\gamma=1$ corresponds 
 to a model in which resetting particles travel at the same velocity as the running particles (the only difference is the absence of tumbles during the resetting to the origin).\\

The kinetic equations in the steady state at $x > 0$ read:
\begin{equation}\label{sysNonInst}
\begin{split}
0 &= -\frac{d}{dx}\left( \rho_+ v(\rho_{\text{tot}}) \right) - \left(r + \frac{\alpha}{2}\right) \rho_+ + \frac{\alpha}{2}\rho_-,\\
0 &= \frac{d}{dx}\left( \rho_- v(\rho_{\text{tot}}) \right) - \left(r + \frac{\alpha}{2}\right) \rho_- + \frac{\alpha}{2}\rho_+,\\
0 &= \frac{d}{dx}\left( \rho_r v_r(\rho_{\text{tot}}) \right) + r (\rho_+ + \rho_-).
\end{split}
\end{equation}

 The fluxes $R$ and $S$ are modified as follows in the notations of the model. Moreover, we introduce the additional resetting flux $J(x)$:
\begin{equation}
    \begin{split}
R(x) :=& v(\rho_{tot}(x)) \rho(x),\\
S(x) :=& v(\rho_{tot}(x)) \sigma(x), \qquad {\mathrm{with}}\qquad \sigma(x) =\rho_+(x) - \rho_-(x),\\
J(x) :=& v_r(\rho_{tot}(x)) \rho_r(x).
    \end{split}
\end{equation}
 In the limit of instantaneous resetting, the density of resetting particles $\rho_r$ goes to zero, and $\rho$ becomes equal to $\rho_{\text{tot}}$.  The expression of the fluxes $R$ and $S$ given in Eq. \eqref{defRS} is recovered.\\

 The combination $R^2 - \frac{r}{r+\alpha}S^2$ is constant in each phase, as in the case of ordinary resetting.
 In the gas phase, all densities go to zero at both infinities, hence this constant is zero, and the same proportionality relation holds as in the case of instantaneous resetting:
\begin{equation}
 \sigma(x) = \sqrt\frac{r}{r+\alpha} \rho(x).
\end{equation}
 Moreover, the density $\rho$ of movers can be expressed as a space derivative, by taking the sum of the first two equations in Eq. \eqref{sysNonInst}:
 \begin{equation}\label{evolS}
 r\rho = - \frac{\partial S}{\partial x}. 
 \end{equation}
 Hence,
 \begin{equation}
  r\rho = -\sqrt{\frac{r}{r+\alpha}} \frac{\partial R}{\partial x}. 
 \end{equation}
On the other hand, the quantity $r\rho$ is also equal to $J'(x)$.  Substituting,
\begin{equation}
\frac{d}{dx}\left( J(x)  - \sqrt{\frac{r}{r+\alpha}}R(x)  \right) =0.
\end{equation}
The quantity $J(x)  - \sqrt{\frac{r}{r+\alpha}}R(x)$ is therefore a constant in each phase.  
The densities $\rho$ and $\rho_r$  go to zero at infinity. This constant is therefore zero in the gas. Hence,  
\begin{equation}\label{proprhorrrho}
\rho_r = \frac{1}{\gamma}\sqrt{\frac{r}{r+\alpha}} \rho.
\end{equation}
 In the gas phase, the total density is therefore $\rho_{\text{tot}} = \rho_r + \rho = \rho\left(1 + \frac{1}{\gamma}\sqrt{\frac{r}{r+\alpha}}\right)$, and Eq. \eqref{evolS} becomes
\begin{equation}\label{evolSPrac}
\begin{split}
\frac{d}{d x}\left[  v_0\left(1- \frac{\rho}{\rho_m}\left(1 + \frac{1}{\gamma}\sqrt{\frac{r}{r+\alpha}}\right)\right)\sqrt{\frac{r}{r+\alpha}}\rho\right] =& -r\rho.\\
\end{split}
\end{equation}
We obtain the analogue of Eq. \eqref{eq:separable}
\begin{equation}\label{eq:xiDef}
\begin{split}
\frac{d}{d x}\left[  \rho - \frac{\xi}{\rho_m}\rho^2\right] =& -\chi \rho, \qquad{\mathrm{with}}\qquad \xi = 1 + \frac{1}{\gamma}\sqrt{\frac{r}{r+\alpha}}.\\
\end{split}
\end{equation}
In the gas phase, the density $\rho$ of movers therefore satisfies the separable equation
\begin{equation}\label{evolSPrac}
\begin{split}
\frac{\rho'}{\rho} - \frac{2\xi}{\rho_m} \rho' =& -\chi.
\end{split}
\end{equation}
 The solution proceeds using the same technical steps as in Sections \ref{sec:droplet} and \ref{sec:gas}. The calculations are detailed in 
  Appendix \ref{repeat}. The width of the droplet is identical to the one we found in Section \ref{sec:droplet}.
 The results are summarized in Table \ref{tab:summary_profiles}, which recovers the results of the previous sections by sending the factor $\gamma$ to $\infty$ (or equivalently by substituting $1$ to the constant $\xi$).

  \begin{table}[htbp]
    \centering
    \renewcommand{\arraystretch}{1.8}
    \begin{tabular}{clc}
        \toprule
        \textbf{Quantity} & \textbf{Expression} & \textbf{Condition of existence} \\
        \midrule
        \multicolumn{3}{l}{\textbf{Two-phase steady state (Droplet + Gas)}} \\
        \midrule
        $\rho(x)$ & 
        $\left[\frac{\rho_m}{\xi} e^{-\rho_m \chi (|x|-b)}\right] \mathbbm{1}( |x|\leq b) 
        -\frac{\rho_m}{2\xi} W_0\left( -C_b e^{-C_b - \chi(|x|-b)}\right) \mathbbm{1}( |x|> b)$  & 
        \multirow{5}{*}{$\displaystyle \rho_m \geq 4 \quad \text{and} \quad M\chi > 2$} \\
        $\sigma(x)$ & $\displaystyle \sqrt{\frac{r}{r+\alpha}} \rho(x)$ & \\
        $\rho_+(x)$ & $\displaystyle \frac{1}{2}\left(1 + \sqrt{\frac{r}{r+\alpha}}\right) \rho(x)$ & \\
        $\rho_-(x)$ & $\displaystyle \frac{1}{2}\left(1 - \sqrt{\frac{r}{r+\alpha}}\right) \rho(x)$ & \\
        $\rho_r(x)$ & $\displaystyle (\xi - 1) \rho(x)$ & \\
        \midrule
        \multicolumn{3}{l}{\textbf{Single-phase steady state (Gas)}} \\
        \midrule
        $\rho(x)$ & $\displaystyle -\frac{\rho_m}{2\xi} W_0\left( -C_0 e^{-C_0 - \chi |x|} \right)$ & \multirow{5}{*}{$\displaystyle M\chi \leq \frac{\rho_m}{2}$} \\
        $\sigma(x)$ & $\displaystyle \sqrt{\frac{r}{r+\alpha}} \rho(x)$ & \\
        $\rho_+(x)$ & $\displaystyle \frac{1}{2}\left(1 + \sqrt{\frac{r}{r+\alpha}}\right) \rho(x)$ & \\
        $\rho_-(x)$ & $\displaystyle \frac{1}{2}\left(1 - \sqrt{\frac{r}{r+\alpha}}\right) \rho(x)$ & \\
        $\rho_r(x)$ & $\displaystyle (\xi - 1) \rho(x)$ & \\
        \bottomrule
    \end{tabular}
    \caption{{\bf{Stationary spatial profiles for the extended model with finite resetting velocity.}} The  parameters are the inverse length scale $\chi = \frac{1}{v_0}\sqrt{r(r+\alpha)}$ and the droplet half-width $b = \frac{1}{\rho_m\chi}\ln\left(\frac{M\chi}{2}\right)$. We have defined the dimensionless constants $C_b = 1 - \sqrt{1 - \frac{4}{\rho_m}}$ and $C_0 = 1 - \sqrt{1 - \frac{2M\chi}{\rho_m}}$. We used the definition $\xi=1 + \frac{1}{\gamma}\sqrt{\frac{r}{r+\alpha}}$, where $\gamma v_0$ is the maximum velocity of resetting particles. The parameter $\xi= 1 + \frac{1}{\gamma}\sqrt{\frac{r}{r+\alpha}}$ was defined in Eq. \eqref{eq:xiDef}. The instantaneous resetting model is exactly recovered by setting the velocity parameter $\xi$ to $1$ (and thus $\rho_r = 0$). If the resetting particles travel at the same velocity as the running particles, $\xi$ becomes $\xi= 1 + \sqrt{\frac{r}{r+\alpha}}$.}
    \label{tab:summary_profiles}
\end{table}



\section{Discussion and Conclusion}
\label{sec:conclusion}

We have investigated the non-equilibrium stationary states of a one-dimensional system of run-and-tumble particles undergoing motility-induced phase separation subject to local stochastic resetting at the origin. We demonstrated that above a critical resetting rate $r_c$, a stable two-phase structure emerges, characterized by a dense liquid droplet centered at the origin coexisting with a surrounding supersaturated gas phase. Through exact analytical formulations, we characterized the stationary density profile.\\

There is some  analogy between the droplet size in our framework (arguably the simplest model for phase separation under local resetting) and the optimization of the mean first-passage time in Brownian search with resetting \cite{evans2011optimal}. Although they represent fundamentally different observables, both exhibit an exact, analytically calculable optimum resulting from the competition between two distinct hindering regimes. In the strong-resetting limit, both systems become overly localized: the Brownian searcher is continually pulled back to its starting point before reaching its target, while a localized liquid droplet asymptotically shrinks to a dense pack of RTPs alternating between short ballistic trajectories and resetting to the origin. Conversely, in the weak-resetting regime, the MFPT diverges, while a system of RTPs fails to maintain a macroscopic condensed phase. However, this analogy remains strictly partial: whereas the MFPT typically diverges smoothly as the resetting rate vanishes, the droplet width in our model does not merely decay to zero but undergoes a phase transition, remaining exactly zero below the finite critical rate $r_c$.\\

Generalizing this framework to two and three dimensions would make it more relevant to experimentally realized systems. Promising extensions include investigating higher-dimensional active droplets where surface tension and transverse fluctuations play a role \cite{santra2020run, smith2022exact}.
Interface fluctuations and geometric constraints alter both MIPS dynamics and resetting kinetics. The solution to the equation of motion of an RTP in two dimensions is known \cite{santra2020run}. It can be adapted to the evolution equation of the density in the fluid phase, but the non-linear equation in the gas phase contains a non-linear term giving rise to a hierarchy of equations. More realistic models of resetting dynamics could involve non-instantaneous tumbling, non-Poissonian resetting time distributions or spatially extended resetting regions \cite{grange2026first,dhar2019run, basu2020exact, gueneau2024optimal}.\\


\begin{appendices}

\section{Discrete formulation for numerical simulation}
\label{app:discrete}

 For  numerical implementation, let us formulate a discrete version of the model.  There is an integer number of particles with any given velocity state at each site. The distance between two consecutive sites is $a$. For every integer $l$, there is a site at position $la$ (we say that this site is {\emph{labelled}} $l$).  The model is a zero-range process in each of the two populations of left-movers and right-movers, with resetting at rate $r$ and tumble events that couple left-movers and right-movers.\\

 Each particle has a mass $m>0$. Consider a site labelled $l$, occupied by $n$ particles. 
 In our system of units, a mass is a length, so the density at the site labeled $l$ is the total length of the particles occupying the site, divided by the length $a$:
 \begin{equation}\label{eq:disreteDensity}
     \rho( la ) = \frac{n m}{a},\qquad (l\in \mathbbm{Z}).
 \end{equation}


 Let us denote by $p_l^+(n)$ (resp. $p_l^-(n)$) the probability of presence of $n$ particles with positive velocity (resp. negative velocity) at site labeled $l$, in the steady state. Let us denote by $\overline{n}_l^+$ (resp. $\overline{n}_l^-$) the mean number of particles with positive (resp. negative) velocity at site labelled $l$ in the steady.   Assuming a factorized steady state, 
\begin{equation}
\overline{n}_l^\pm = \sum_{n\geq 0 } n p_l^\pm(n).
\end{equation}
 The zero-range dynamics implies
\begin{equation}\label{stationarityEqDiscrete}
\begin{split}
0 =& -(r+\lambda_l) n p_l^+(n) + \lambda_{l-1} \overline{n}_{l-1}^+ p_l^+(n-1) + \frac{\alpha}{2} \overline{n}_{l} p_l^+(n-1),\\ 
0 =&  -(r+\lambda_l) n p_l^-(n) + \lambda_{l+1} \overline{n}_{l+1}^- p_l^-(n-1) + \frac{\alpha}{2} \overline{n}_{l}^+ p_l^-(n-1), 
\end{split}   
\end{equation}
 where $\lambda_l$ is a phase-dependent rate chosen so that $\lambda_l a$ reproduces the value of the velocity of particles in the continuum model: 
\begin{equation}
\lambda_l = \frac{v}{a},   
\end{equation}
  The phase is liquid if the number $n$ at the site is greater than the critical number $n_m$ such that the density at the site equals the critical value $\rho_m$:
\begin{equation}
 \rho_m = n_m \frac{m}{a}. 
\end{equation}
 Hence, as a site labelled $l$ occupied by $n$ particles:
\begin{equation}
\lambda_l =  \frac{v_0}{a}\left( 1 - \frac{n m}{\rho_m a}  \right) \mathbbm{1}\left( n < \frac{\rho_m a}{m}  \right) +\frac{v_0}{\rho_m a} \mathbbm{1}\left( n \geq \frac{\rho_m a}{m}  \right) .
\end{equation}

The probability distribution $p_l^\pm(.,t)$ is therefore a Poisson distribution (this is a particular case of Eq. (11) in \cite{thompson2011lattice}). Denoting its mean by $\overline{n}^\pm_{l}$, we obtain
\begin{equation}
p_l^\pm(n) = e^{- \overline{n}^\pm_{l}} \frac{ \left(  \overline{n}^\pm_{l}\right)^n}{n!}\qquad (n\in\mathbbm{N}).
\end{equation}
Summing the above equations over $n$ yields the following system of equations satisfied by the mean numbers of particles:
 \begin{equation}\label{stationarityEqPoisson}
\begin{split}
0 =& -\left(r+\frac{\alpha}{2}+\lambda_l\right) \overline{n}^+_{l}+ + \lambda_{l-1} \overline{n}_{l-1}^+  + \frac{\alpha}{2} \overline{n}_{l}^-,\\ 
0 =&  -\left(r+\frac{\alpha}{2}+\lambda_l\right) \overline{n}^-_{l} + \lambda_{l+1} \overline{n}_{l+1}^- + \frac{\alpha}{2} \overline{n}_{l}^+. 
\end{split}   
\end{equation}




To initialize the model in a configuration corresponding to a supersaturated gas, we need a discrete analogue of the exponential term $e^{-\chi x}$  that appears in the expressions of the density of the gas in the continuum model in each of the phases (as $e^{-\rho_m\chi x}$ in the liquid phase and in the argument of the Lambert function in the gas phase). Let us make the following Ansatz for the mean number of particles at site $l$ in the steady state in the liquid phase:
\begin{equation}
\overline{n}_{l}^\pm = \overline{n}_{0}^\pm \beta^{|l|}.
\end{equation}

Hence,
\begin{equation}
\begin{split}
0 =& -\left(r+\frac{\alpha}{2}+\lambda_l\right) \overline{n}^+_{0}\beta^l + \lambda_{l-1} \overline{n}^+_{0}\beta^{l-1}  + \frac{\alpha}{2} \overline{n}^-_{0}\beta^l,\\ 
0 =&  -\left(r+\frac{\alpha}{2}+\lambda_l\right) \overline{n}^-_{0}\beta^l  + \lambda_{l+1} \overline{n}^-_{0}\beta^{l+1}  + \frac{\alpha}{2} \overline{n}^+_{0}\beta^l,
\end{split}   
\end{equation}
with 
\begin{equation}
    \lambda_l = \frac{v_0}{\rho_m}, \qquad(l\in \mathbbm{Z}).
\end{equation}

Simplifying by $\beta^l$, this system reads as the image of the vector $(\overline{n}^+_{0},\overline{n}^-_{0})^T$ by a matrix. The determinant of this matrix must be zero for the system to have non-zero density:
\begin{equation}\label{sysnn}
\begin{pmatrix}
\lambda\beta^{-1} - (r+\frac{\alpha}{2} + \lambda) & \frac{\alpha}{2}\\
\frac{\alpha}{2} & \lambda\beta - (r+\frac{\alpha}{2} + \lambda)\\
\end{pmatrix}
\begin{pmatrix}
\overline{n}^+_{0}\\
\overline{n}^-_{0}
\end{pmatrix}
= 
\begin{pmatrix}
0\\
0
\end{pmatrix}.
\end{equation}
For this equation to have a non-trivial solution, the determinant of the matrix must be zero:
\begin{equation}
[\lambda\beta^{-1} - (r+\frac{\alpha}{2} + \lambda)][ \lambda\beta - (r+\frac{\alpha}{2} + \lambda) ] -\frac{\alpha^2}{4} =0.
\end{equation}
The quantity $\beta$ therefore satisfies a quadratic equation
\begin{equation}
\beta^2 -\frac{ \lambda^2 -\frac{\alpha^2}{4} +(r+\frac{\alpha}{2} + \lambda)^2}{\lambda\left( r+\frac{\alpha}{2} + \lambda\right)}\beta +1=0,\\
\end{equation}
The product of the two roots equals $1$. The relevant value of $\beta$ (ensuring that the mean number of particles is a decreasing function of the distance to the origin) is therefore the minimum of the two roots:
\begin{equation}\label{betaExpr}
\beta = \frac{1}{2}\left[  \frac{ \lambda^2 -\frac{\alpha^2}{4} +(r+\frac{\alpha}{2} + \lambda)^2}{\lambda\left( r+\frac{\alpha}{2} + \lambda\right)} - \sqrt{\left(\frac{ \lambda^2 -\frac{\alpha}{2}^2 +(r+\frac{\alpha}{2} + \lambda)^2}{\lambda\left(r+\frac{\alpha}{2} + \lambda\right)}\right)^2 - 4  } \right].
\end{equation}
Taylor expansion yields, in the limit of short step size $a$,
\begin{equation}
\beta = 1 - \rho_m \chi a +o(a).
\end{equation}
The spatial dependence becomes
\begin{equation}
 \beta^{|l|} \simeq \left( 1- \frac{\chi}{\rho_m}a\right)^{\frac{|x|}{a}} \simeq \exp\left( \frac{|x|}{a}\log\left(  1- \rho_m \chi a \right)\right)\simeq e^{-\rho_m\chi |x|},
\end{equation}
 which is consistent with the continuum model.\\

The evolution of the system is simulated as follows:\\
{\bf{1. Choice of parameters.}} Pick the values of the maximum velocity $v_0$, flipping rate $\alpha$, density parameter $\rho_m$, resetting rate $r$, total mass $M$, particle mass $m$. Pick the size of the observed lattice, say $K\chi^{-1}$. Calculate the lattice spacing $a$ by demanding that $K\chi^{-1} = (2L+1)a$, where $L$ is a large integer, and $K$ is a moderate integer. Calculate $\beta$ using Eq. \eqref{betaExpr}.\\ 
{\bf{2. Initialization of configuration.}} For each $l$ in $[-L..L]$, calculate $\overline{n}_l$ and draw a Poisson variable of mean $\overline{n}_l$, which is the total number of particles at site $l$. The signs of the velocities of each of these particles are drawn independently from $\{-1,+1\}$. The probability of drawing a positive is $\frac{1}{2}\left( 1 + \sqrt{\frac{r}{r+\alpha}}\right)$ because $\rho_\pm(x) = \frac{1}{2}\left( 1 + \sqrt{\frac{r}{r+\alpha}}\right)\rho(x)$  (see Table \ref{tab:summary_profiles}).  
The absolute value of the velocity of the particles is $v(\rho)$, where $\rho$ is the local density evaluated using Eq. \eqref{eq:disreteDensity}.\\   
{\bf{2. Gillespie algorithm.}} at each step, the total resetting, jumping and flipping rates are evaluated. If $C$ is the total number of particles, these are $R := Cr$, $F:=C\alpha/2$ and $J:=\sum_{k=1}^C a^{-1}v_k$ respectively, with $v_k$ the  velocity evaluated in terms of the density at the site occupied by the $k-th$ particle.\\
A rate $\tau$ is uniformly picked between $0$ and the total rate $R+F+J$. If $\tau$ is in $[0,R[$ the next event is a reset, a particle is  picked uniformly from the system and reset to the origin. If $\tau$ is in $[R,R+F[$ the next event is a flip, a particle is picked uniformly from the system and the sign of its velocity is flipped. Otherwise, the rate is an interval of the form $[\sum_{k=1}^m a^{-1}v_k, \sum_{k=1}^{m+1} a^{-1} v_k[$. The $m$-th particle jump to 
 the right-neighboring (left left-neighboring) site if its velocity is positive (resp. negative).\\





\section{Phases under non-instantaneous resetting}\label{repeat}

\subsection*{Steady state with a droplet centered at the origin}
 Let us look for a steady state with a liquid droplet occupying the domain $[-b,b]$ (for unknown $b$), surrounded by a gas phase.  Integrating Eq. \eqref{evolSPrac} between the boundary $b$ and some position $x>b$ in the gas,
\begin{equation}
  \ln\left(\frac{\rho(x)}{\rho(b^+)}\right)  -\frac{2\xi}{\rho_m} \left(\rho(x) - \rho(b^+) \right)= -\chi( x-b).
\end{equation}
Exponentiating and rearranging,
\begin{equation}\label{gasFinite}
\rho(x) = -\frac{\rho_m}{2\xi} W_0\left(-\frac{2\xi\rho(b^+) }{\rho_m}e^{-\frac{2\xi\rho(b^+) }{\rho_m}} e^{-\chi(x-b)} \right), \qquad (x>b).
\end{equation}
On the other hand, the evolution equation in the liquid phase 
  (where $\rho_{\text{tot}}\geq \rho_m$) is Eq. \eqref{eq:evolEqLiquid}, as in the model with instantaneous resetting. By the same reasoning as in the case of instantaneous resetting,
 \begin{equation}
\sigma(x) = \sqrt{\frac{r}{r+\alpha}}\rho(x) = \frac{r}{v_0\chi}\rho(x), 
 \end{equation}
which is the proportionality relation holding in the gas. The relation $\rho_r = \frac{1}{\gamma} \sqrt{\frac{r}{r+\alpha}}\rho$ therefore holds in the entire space, and
\begin{equation}
  \rho_{\text{tot}}(x) =  \rho_r(x) + \rho(x) = \xi \rho(x),\qquad (x\in\mathbbm{R}).
\end{equation}
Hence the density profile in the interval $[0,b]$:
\begin{equation}
\begin{split}
 \rho(x) =& \frac{\rho_m}{\xi} e^{-\rho_m\chi x} \\
 \sigma(x) =&  \frac{\rho_m}{\xi} \sqrt{\frac{r}{r+\alpha}} e^{-\rho_m\chi x}   \qquad (x\in[0,b]).
\end{split}    
\end{equation}

At the interface between the two phases, the total density of particles in the liquid equals its minimum possible value:  
\begin{equation}
\begin{split}
\rho_m =& \rho_{tot}( b^-).\\
\end{split}
\end{equation}
On the other hand, we can express the total density in the liquid using Eq. \eqref{proprhorrrho}:
\begin{equation}
\rho_{tot}(b^-)= \left( 1 + \frac{1}{\gamma}\sqrt{\frac{r}{r+\alpha}} \right)\rho(b^-),
\end{equation}
Hence the boundary condition on the density of movers $\rho$ in the liquid:
\begin{equation}
\rho(b^-) = \frac{\rho_m}{\xi}.
\end{equation}

Let us impose the continuity of the current $S$ across the interface:
\begin{equation}
 \frac{v_0}{\rho_m}\sigma( b^-) = v_0\left( 1 - \xi\frac{\rho(b^+)}{\rho_m}\right) \sigma( b^+),\\
\end{equation}
As the current $\sigma$ is proportional to the density of movers $\rho$, we obtain a quadratic equation in $\xi\rho(b^+)$:
\begin{equation}
\begin{split}
 \frac{v_0}{\rho_m}  \frac{r}{v_0 \chi}\frac{\rho_m}{\xi} =&  v_0\left( 1 - \xi\frac{\rho(b^+)}{\rho_m}\right)\frac{r}{v_0\chi}\rho(b^+),\\
  1 =& 
 \left( 1 - \xi\frac{\rho(b^+)}{\rho_m}\right) \xi\rho(b^+),\\
 \left(\xi \rho(b^+)\right)^2 -\rho_m \xi \rho( b^+) + \rho_m =&0,\\
\end{split}
\end{equation}
The solution with a droplet exists only if
\begin{equation}
\rho_m \geq 4,
\end{equation}
and reads
\begin{equation}
\xi\rho(b^+) = \frac{\rho_m}{2}\left(  1- \sqrt{1 - \frac{4}{\rho_m} }\right).
\end{equation}
Substituting into Eq. \eqref{gasFinite} yields the density of movers in the gas
\begin{equation}
\rho(x) = -\frac{\rho_m}{2\xi} W_0\left( - \left(1-\sqrt{1-\frac{4}{\rho_m}}\right) e^{-\left(1-\sqrt{1-\frac{4}{\rho_m}}\right)} e^{-\chi(x-b)} \right),\qquad (x>b).
\end{equation}
 The density profile in the gas occupying the region $]-\infty,-b[$ is obtained by substituting $|x| - b$ to $x-b$ in the above expression.
 This is the density profile reported in Table \ref{tab:summary_profiles}.\\

The proportionality relation between $\rho_r$ and $\rho$ is the same in the liquid and in the gas.  
Conservation of mass is expressed as
\begin{equation}
\begin{split}
  \frac{M}{2} =& \int_0^\infty( \rho(x) + \rho_r(x))dx\\
  =&(1+(\xi-1)) \int_0^\infty \rho(x) dx\\
  =& \xi\left[ \int_0^b \rho( b_-) e^{-\rho_m \chi ( x - b)} dx -\frac{\rho_m}{2\xi} \int_b^\infty W_0\left( - \left(1-\sqrt{1-\frac{4}{\rho_m}}\right) e^{-\left(1-\sqrt{1-\frac{4}{\rho_m}}\right)} e^{-\chi(x-b)} \right) dx \right]\\
  =& \rho_m\int_0^b \rho( b_-) e^{-\rho_m \chi ( x - b)} dx -\frac{\rho_m}{2} \int_b^\infty W_0\left( - \left(1-\sqrt{1-\frac{4}{\rho_m}}\right) e^{-\left(1-\sqrt{1-\frac{4}{\rho_m}}\right)} e^{-\chi(x-b)} \right) dx.  
\end{split}
\end{equation}
 Mass conservation is therefore expressed by the same condition  as in Eq. \eqref{eq:mass_conservation_integral}. 
  The value of the length $b$ is therefore independent of the factor $\gamma$:
\begin{equation}
    b = \frac{1}{\rho_m \chi }\log\left( \frac{M\chi}{2}\right).
\end{equation}

\subsection*{Steady state in gas phase}
 Let us look for a steady state with a single (gas) phase. Integrating Eq. \eqref{evolSPrac} between the origin and position $x>0$ 
\begin{equation}
  \ln\left(\frac{\rho(x)}{\rho(0)}\right)  -\frac{2\xi}{\rho_m} \left(\rho(x) - \rho(0) \right)= -\chi x.
\end{equation}
Exponentiating and rearranging,
\begin{equation}
\rho(x) = -\frac{\rho_m}{2\xi} W_0\left(-\frac{2\xi\rho(0) }{\rho_m}e^{-\frac{2\xi\rho(0) }{\rho_m}} e^{-\chi x} \right).
\end{equation}
Mass conservation is expressed as
\begin{equation}
 \frac{M}{2} = \int_0^\infty \left(\rho( x) + \rho_r(x) \right) dx   
\end{equation}
The density of resetting particles is  proportional to the density of running particles according to Eq. \eqref{proprhorrrho}. Hence,
\begin{equation}
 \begin{split}
 \frac{M}{2} =& \int_0^\infty (\rho( x) + (\xi - 1) \rho(x) ) dx\\
 =& \int_0^\infty -\frac{\rho_m}{2} W_0\left(-\frac{2\xi\rho(0) }{\rho_m}e^{-\frac{2\xi\rho(0) }{\rho_m}} e^{-\chi x} \right)dx.
 \end{split}
\end{equation}
 The quantity $\xi\rho(0)$ therefore satisfies the same equation (Eq. \eqref{peakDensity}) that  is satisfied by $\rho(0)$ in the model with instantaneous resetting:
\begin{equation}
    \frac{\left(\xi \rho(0)\right)^2}{\rho_m} - \xi \rho(0) + \frac{M\chi}{2} = 0. 
\end{equation}
 The condition of existence of the single-phase solution is the same as in the ordinary case. Assuming the condition
 \begin{equation}
 M\chi \leq \frac{\rho_m}{2}.
 \end{equation}
 is satisfied, we obtain  

\begin{equation}
  \rho(0) = \frac{\rho_m}{2\xi}\left(1-\sqrt{1- \frac{2M\chi}{\rho_m}} \right).
\end{equation}
 The density of running particles is expressed in the entire space as
\begin{equation}
\begin{split}
 \rho(x) =& -\frac{\rho_m}{2\xi} W_0\left(\left(-1+\sqrt{1- \frac{2M\chi}{\rho_m}} \right)e^{\left(-1+\sqrt{1- \frac{2M\chi}{\rho_m}} \right) -\chi |x|} \right).\\  
\end{split}  
\end{equation}
This is the density profile reported in the second part of Table \ref{tab:summary_profiles}.

\end{appendices}

\bibliographystyle{ieeetr}
\bibliography{Bibref}

\end{document}